**Developing Digital Twins for Earth Systems: Purpose, Requisites, and Benefits**

Yuhan Rao[1,2], Rob Redmon[2], Kirstine Dale[3], Sue E. Haupt[4], Aaron Hopkinson[3], Ann Bostrom[5], Sid Boukabara[6], Thomas Geenen[7], David Hall[8], Benjamin D. Smith[9], Dev Niyogi[10], V. Ramaswamy[11], Eric A. Kihn[2]

1. Cooperative Institute for Satellite Earth System Studies, North Carolina State University
2. National Centers for Environmental Information, NOAA
3. UK Met Office
4. National Center for Atmospheric Research
5. University of Washington, Seattle
6. Earth Science Division, NASA Science Mission Directorate
7. European Centre for Medium-Range Weather Forecasts
8. NVIDIA
9. Jet Propulsion Laboratory, NASA
10. University of Texas, Austin
11. Geophysical Fluid Dynamics Laboratory, NOAA

**Abstract:** The accelerated change in our planet due to human activities has led to grand societal challenges including health crises, intensified extreme weather events, food security, environmental injustice, etc. Digital twin systems combined with emerging technologies such as artificial intelligence and edge computing provide opportunities to support planning and decision making to address these challenges. Digital twins for Earth systems (DT4ESs) are defined as the digital representation of the complex integrated Earth system including both natural processes and human activities. They have the potential to enable a diverse range of users to explore what-if scenarios across spatial and temporal scales to improve our understanding, prediction, mitigation, and adaptation to grand societal challenges. The 4th NOAA AI Workshop convened around 100 members who are developing or interested in participating in the development of DT4ES to discuss a shared community vision and path forward on fostering a future ecosystem of interoperable DT4ES. This paper summarizes the workshop discussions around DT4ES. We first defined the foundational features of a viable digital twins for Earth system that can be used to guide the development of various use cases of DT4ES. Finally, we made practical recommendations for the community on different aspects of collaboration in order to enable a future ecosystem of interoperable DT4ES, including equity-centered use case development, community-driven investigation of interoperability for DT4ES, trust-oriented co-development, and developing a community of practice.

## 1. Current Development of Digital Twins for Earth Systems

Due to the accelerating rate of change in our planet, we humans find ourselves facing global and regional crises. These include health, ecosystem, and socioeconomic crises with ongoing and predicted impacts to coastal community resilience, sustainable fisheries, air quality and heat exposure in historically marginalized communities, etc ([IPCC, 2023](#)). Digital twin technology, combined with emerging technologies such as artificial intelligence and edge computing, is seen as an opportunity to help us face these challenges by bringing appropriate information into the hands of decision-makers and the various socioeconomic actors quickly, more efficiently, and at the right spatial and temporal resolutions.

In our view, the Earth system is an integrated one that includes both natural processes (e.g., weather and climate phenomena, ecosystem processes) and human activities (e.g., economic development, land use). To be realistic, an ecosystem of interoperable digital twins of Earth systems is needed to address the challenges that society collectively faces. These digital systems can enable users to rapidly run hypothetical *what-if* scenarios across spatial and temporal scales to improve our understanding, prediction, mitigation, and adaptation to grand societal challenges ([Nativi et al., 2021](#)). These digital twin systems are needed in order to provide users with tailored access to high-quality observations and predicted knowledge along with associated uncertainty information for user-specific scenario development for decision support. Equitable access to this information can improve equitable responses to extreme weather and natural hazards, mitigate extreme climate-related risks, address global change challenges, and meet Sustainable Development Goals (e.g. [Nativi et al., 2021](#); [Tzachor et al., 2022](#)). To achieve these capabilities, Artificial Intelligence / Machine Learning applications will provide critical value throughout digital twin systems and supporting workflows, including fusing multi-platform observations, accelerating large scale simulations, enabling *what-if* workflows, tailoring user community risk communication, etc (e.g., [Pathk et al., 2022](#), [ECMWF, 2023](#)).

For the purposes of this white paper, we define a Digital twin for Earth system (DT4ES) as the digital representation of the complex Earth system that allows us to visualize, monitor, and forecast natural and human activities on our planet whose external and internal component interfaces adhere to open standards to enable an overall ecosystem of connected digital twins. Typically, digital twin systems form part of a wider cyber-physical infrastructure composed of the following components as illustrated in Figure 1:

a) Digital infrastructure to host and implement digital twin systems which includes data lakes, computation models, associated tools and software, and hosting interface;
b) Information management framework, which includes data and information standards and protocols that enable the transmission of information across different components of the digital twin system;
c) Data integration system that contains real world observations and model simulations enabling the linkage between the digital representation and the Earth system. To enable integrated modeling and decision making, these data should represent both natural and human processes that are relevant to the use cases.
d) Computational model that captures the pertinent natural and human processes of the integrated Earth system;
e) User interface that enables diverse users to interact with digital twin systems and empower decision making for a wide range of use cases.

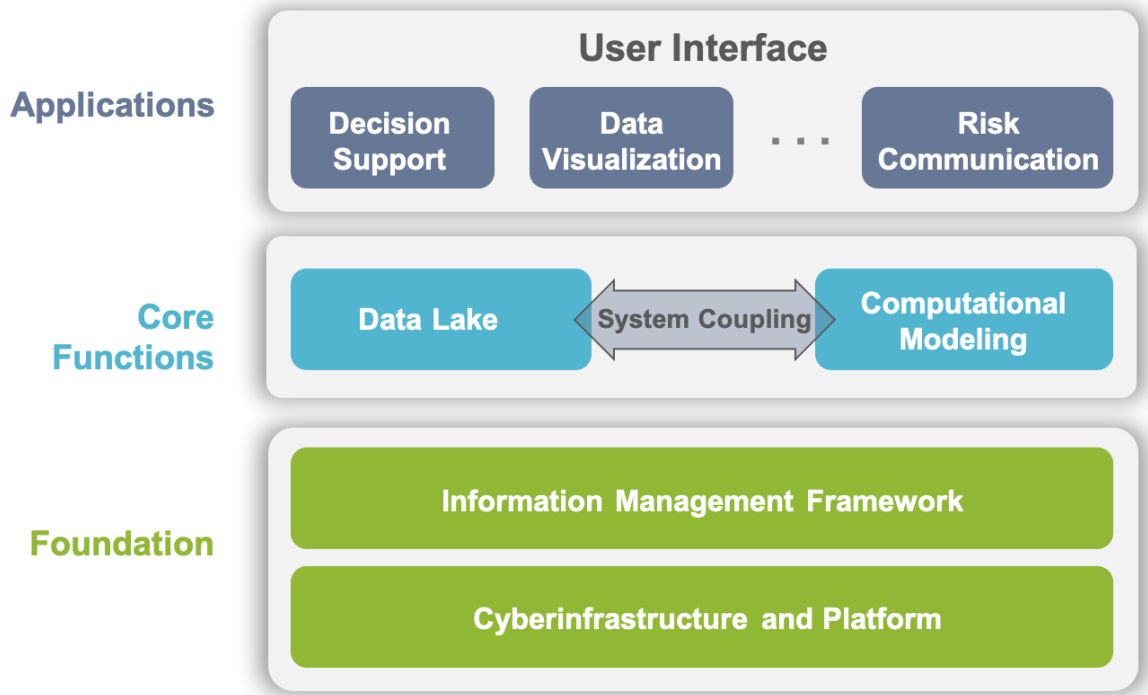

**Figure 1**: Representative conceptualization of a DT4ES architecture.

Several institutions, including NOAA, NASA, ECMWF, the UK Met Office, academic institutions, and private industry are developing various DT4ES implementations. As a community we recognize the need to develop a shared vision and international open standards to ensure that digital twins are interoperable and easily integrated/federated into other digital twin systems for different socioeconomic sectors. The 4th NOAA AI Workshop in 2022 provided a platform for further building an internationally diverse community of practice on interoperable digital twins through a series of talks, panel conversations and interactive discussions covering shared community vision forming, digital twin components, roles for AI, and equity and ethical considerations for different types of users.

There are various national and international initiatives being spun up that aim to develop digital twins for Earth systems. Some are developed by governments or intergovernmental entities, and others by the private sector. We list a few of these in the next paragraphs.

The United Nations Decade of Ocean Sciences for Sustainable Development is coordinating an international effort to develop a [Digital Twin of the Ocean (DITTO)](). The effort aims to advance a common understanding, establish recommended practices, and advance a shared digital framework for the development of future digital twins of the ocean. DITTO's vision is to use digital twin systems to support ocean science, conservation, governance, and a sustainable ocean economy. Through engagement and pilot studies, DITTO promotes co-design of digital twin systems with targeted end users, raises awareness of their uses and applications, and demonstrates their potential for decision making across multiple sectors including ocean governance.

Last updated: 2023-06-16

In 2020, the European Commission established a flagship initiative – [Destination Earth (DestinE)](#) – to support Europe's Green Deal and Digital Strategy. The objective of DestinE is to develop a highly accurate digital model of the Earth system globally to monitor, simulate, and predict the outcome of integrated Earth and human systems. This digital model will support Europe's decision making to mitigate and adapt to climate change and extreme events and prepare for their socio-economic impact for a resilient and sustainable future. The DestinE is a large-scale initiative with an initial commitment of €150 million from the Digital Europe Programme until mid-2024 to support the first phase implementation involving three major entities: the European Space Agency, EUMETSAT, and the European Centre for Medium-Range Weather Forecasts.

The UK Met Office's vision for digital twins is to transform its "weather and climate national capability to integrate increasing complexity in an ecosystem that supports cyber-physical infrastructure, such as 'environment-aware digital twins'". This means that the Met Office is developing both digital twins that incorporate weather and climate information as well as the capability to facilitate cross-discipline and cross-organisational digital twin development through components that can be integrated within digital twins developed by others so that they are 'environment-aware'. To deliver this vision, the Met Office is implementing a 'roadmap for digital twins' that is informed by the [Information Management Framework for Environmental Digital Twins (IMFe)](#) and advances all of the digital twin components (see Figure 1). Current activities include development of digital twin demonstrators targeted at priority use-cases and the forging of a community across environmental science that can help realize the potential for digital twins to fundamentally transform how we extract value and meaning from models and data.

In addition to the public sector investment, private sectors have also kicked off their development of DT4ES. One of the most high-profile examples is the [Earth-2 Initiative by NVIDIA](#). In 2021, NVIDIA launched its plan to develop digital twins for weather and climate prediction using its Omniverse supercomputer system. Earth-2 aims to improve the skills of extreme weather prediction and climate change projection with the ultimate goal of supporting effective mitigation and adaptation strategies. This is achieved by combining accelerated computing and physics-informed machine learning to provide actionable weather and climate information at regional scales. One important distinction of NVIDIA's initiative is its capability to utilize and design state-of-the-art hardware to support the development of DT4ES.

NOAA, in partnership with industry and sister agencies, is exploring the development of prototype digital twins for Earth observation ([EO-DT](#)) that combines different sources of observations including satellite, ground, and airborne data for various Earth system components (biosphere, cryosphere, atmosphere, ocean, hydrology/land, and space weather). The purpose is to streamline the data acquisition, processing, quality and anomaly monitoring, dissemination, and interfacing with the wide range of users of the observations. This is envisioned to be one of the components of the NOAA DT concept, which focuses on observations fusion and assimilation.

The NASA Earth Science Technology Office recently invested in pilot development of [Earth System Digital Twins (ESDT)](#) in 2021 through its Advanced Information Systems Technology (AIST) program. AIST invested $31 million supporting 28 projects to develop novel technologies that integrate human activities and physical Earth systems. These novel technologies focus on providing comprehensive and consistent digital representation to support the monitoring and prediction of the integrated system. They also focus on developing actionable information, as well as supporting decision making through the exploration of various "what-if" scenarios. All funded projects explore different application areas like public health, agriculture monitoring, water quality and resource management, and wildfire management.

The potential of digital twin technologies also attracted research and development to support sustainable urban development ([Li et al., 2021](#)) and planning, as the United Nations projected that more than 70% of the population will reside in urban areas by 2050. Organizations like the University of Texas, Georgia Institute of Technology, and Argonne National Laboratory are exploring technologies like the Internet of Things (IoT), artificial intelligence, and scalable integrated system modeling to support the development of urban digital twins for future urban planning. In 2022, the World Economic Forum also released a report to highlight a roadmap for urban digital twins to support smart cities in the future ([World Economic Forum, 2022](#)).

## 2. Community Building around DT4ES at the 4th NOAA AI Workshop

In September 2022, NOAA convened its [4th AI Workshop](#) online and included digital twins for Earth systems as its main focus. The goal of this workshop was to bring together participants who are interested in the topic of digital twins across different sectors including academia, government, and industry, and formulate a shared vision to ensure that the future ecosystem of DT4ES can be federated and is interoperable. There were 271 registered participants for the DT4ES theme of the workshop with around 100 members actively participating in the workshop.

The workshop was designed to answer two key questions: 1) how can it be ensured that the bottom-up development of DT4ES across sectors is interoperable? and 2) what are the roles of AI in the interoperable DT4ES ecosystem? To collect broad community input, the workshop activities included lightning presentations from leading research and industry members from NOAA, NASA, ECMWF, Met Office, NVIDIA, Amazon, and academia. These lightning talks spotlighted organizational visions and initiatives around DT4ES, providing a baseline for community members to develop a shared vision for future DT4ES development.

The workshop further explored the major components of DT4ES, including data and observations, modeling and simulation, computing and infrastructure, data analysis and visualization, and user engagement and risk communication (Figure 1). Participants were invited to share their perspectives in different breakout rooms based on research expertise and interests. Here, they discussed the current landscape of DT4ES components as well as community needs for the future DT4ES ecosystem.

During the workshop, participants identified four major gaps and potential pathways forward in developing an interoperable DT4ES ecosystem to address societal challenges.

First, it is necessary to have a common definition for a digital twins for Earth system to support coordination for the development. There are many bottom-up developments across different institutions that all have slightly different interpretations of what a digital twins for Earth systems means. To avoid creating friction that may limit the value of an interoperable DT4ES ecosystem, a common definition is necessary to facilitate and coordinate the development of technologies to support future DT4ES systems.

Second, developing a future DT4ES ecosystem requires strategic coordination. There are emerging developments across sectors internationally on DT4ES. It will be beneficial for all involved organizations to coordinate these efforts and share progress through either existing international coordination groups, or the establishment of a new group to enable the coordination. The coordination should help capture lessons learned from the ongoing development of use cases, facilitate the development of community standards, and formulate long-term plans for the co-development of interoperable DT4ES ecosystems.

Additionally, given the advanced development and adoption of digital twin technologies in other sectors including manufacturing, medical, and engineering industries, participants believe it will be helpful to engage with experts outside of the Earth science domain to learn about their experiences in developing digital twin technologies that are relevant for DT4ES development including their limitations. Lessons on how to address data privacy concerns ([Fuller et al., 2020](#)), ethics implications ([Calzati & van Loenen, 2023](#)), and data-model integration ([Fonseca & Gaspar, 2020](#)) from other domains can be valuable for the Earth system science community. The exchange with cross-industry members will help the Earth science community avoid unnecessary pathfinding and foster co-development of the applicable DT4ES ecosystem that interoperates with existing digital twin technologies and benefits society.

Lastly, maturing the DT4ES ecosystem requires diverse use cases across the spectrum of research to application. As an emerging topic, pilot use case development is crucial to identify the values and limitations of digital twin technologies. Supporting pilot use case development may require broad community commitment from academia, industries, and the public. The development of use cases can inform and be leveraged to create educational content for the broader Earth system science community.

### 3. Defining the foundational features of a viable digital twin for Earth systems

Here we outline the foundational features that a viable DT4ES should have in order to distinguish it from the existing applications and development that may have a similar scope with various aspects of a DT4ES.

First, a viable DT4ES should enable a set of diverse users with differing levels of expertise and needs to interact with the system to explore information under different scenarios that can support relevant decision-making. These decisions can be across a range of domains to support communities to address societal challenges.

Second, a DT4ES should be computationally efficient so as to support users' exploration of different "what if" scenarios. The computational efficiency may be use case dependent, which is similar to current requirements for data latency for different applications. For example, planning

Last updated: 2023-06-16

for rapid response to fast-evolving disasters will have different timeliness requirements from long-range urban planning in response to long-term climate change for a region.

Moreover, a DT4ES should offer a robust representation of both the geophysical and socioeconomic systems at relevant scales for intended use cases. Different from traditional Earth system modeling frameworks, a DT4ES should have the ability to integrate some representation of socioeconomic systems that enable a set of users to explore the outcome of their decisions. A minimal implementation of the integration can be conceptually equated to the overlay between Earth system model simulations with socioeconomic data (e.g., census tract data, population/building density data) to explore the impact of different scenarios on different social sectors. A valuable next step would be to dynamically integrate socioeconomic and natural systems to explore the bi-directional feedback between social sectors on other earth system elements.

Lastly, a DT4ES should provide information that enables users' trust in the system when using it for decision-making. This includes offering a minimum level of verifiability of the system, such as comparison with historical observations and outcomes, and effective information uncertainty that may affect users' decision-making processes.

## 4. Recommendations for the future ecosystem of DT4ES and use case developments

The foundational features proposed in the prior section only pertain to a single Digital Twin System for Earth. However, we agree that currently it is nearly impossible to develop a unified DT4ES that can meet the demand of diverse sectors and use cases. Therefore, we believe the future will be an ecosystem of interconnected Digital Twins for Earth Systems at different scales and for diverse use cases. Our recommendations for the community that is developing the future ecosystem of digital twins for Earth systems are summarized next.

First, a successful ecosystem of DT4ES would be user-centric or application-centric and provides easy access for users of different levels of expertise and with different needs. Equitable access is critical to maximize the societal benefits of DT4ES to address grand challenges such as [sustainable development goals (SDGs)](#) outlined by the United Nations. Also, equitable access does not preclude the development of customizable services and applications using DT4ES to support specific industry needs to benefit economic growth (e.g., agriculture, ocean exploration, retail, and insurance).

Secondly, although interoperability is not mentioned as a foundational feature of a viable DT4ES, it is critical for enabling a functioning ecosystem of DT4ES. Interoperability at different levels (e.g., data, modeling system, user interface) will allow organizations to leverage systems developed by others to reduce development overhead and make efficient research and development investments. To achieve interoperability, there should be a community-accepted standard or set of standards that ensure that key components of DT4ES are interoperable. Efforts have already started to ensure data are interoperable ([Wilkinson et al., 2016](#)) and have been extended to virtual computing environments, research software, and other elements that are critical to DT4ES. These community-driven efforts are fundamental to support the interoperability of DT4ES but might not be sufficient for such an integrated cyber-physical

77

system. Hence, we believe it is beneficial to investigate existing technology standards and conventions and identify key characteristics to ensure the interoperability of DT4ES. For example, Findable, Accessible, Interoperable, Reusable (FAIR) principles have been extended by the community to ensure that software, virtual environments, and other digital objects are interoperable([Chue Hong et al., 2022](#)). Existing data standards and conventions such as the ISO Geographic Information Metadata standard (ISO 19115) and Climate and Forecast Model Metadata Conventions (CF Conventions) can be used to ensure data interoperability. This effort should be done in a coordinated manner involving organizations and stakeholders who are involved in the development of DT4ES.

Thirdly, we acknowledge that a successful ecosystem of DT4ES should follow the principles of co-development to ensure users' trust in the system and minimize potential risks. As artificial intelligence will be a key technology for DT4ES, organizations who plan to develop DT4ES need to address the trustworthiness of AI in order to gain and maintain users' confidence. Research suggests coproduction is critical to the development of trustworthy AI (Stall et al., 2023; McGovern et al., 2022). Co-development requires active and frequent engagement with users throughout the development cycle of AI systems to address user concerns early and properly. It is relatively straightforward to ensure the trustworthiness of an individual DT4ES but it will be more challenging to establish trust in an interconnected ecosystem of DT4ES when information flow from one system to other systems becomes hard to comprehend by users. Measures can be taken to address such issues, for example by establishing benchmarking frameworks to showcase the reliability/validity of DT4ES for various use cases.

Moreover, the Earth science community should learn from other collaborative domains that have developed mature applications of digital twins. DT has been used in other domains such as mechanical engineering, medical applications, and architecture for more than a decade. Developers of DT4ES should leverage lessons learned and mature technologies used in other industries to accelerate the successful development of DT4ES. Ideally, these lessons learned can be collected as a list of shared resources that is accessible to the broader community who are interested in engaging with the development of digital twin systems.

Lastly, the Earth science community should leverage existing forums and create new opportunities when needed to facilitate regular knowledge exchange and community engagement to advance the development of the ecosystem of DT4ES. This could include hosting DT4ES-focused sessions during major scientific conferences such as AGU, EGU, AMS, and IEEE. Additionally, the community should also leverage international organizations and initiatives to discuss the progress of DT4ES development in order to galvanize broad community interest and initiate future activities. When the opportunity matures, the community can form collaborative working groups under existing international groups or initiatives to formalize collaboration and development.

To truly unleash the potential of the technology, a diverse and inclusive community with broad representations of developers, users, domain scientists of both natural and socioeconomic processes, government and international organizations, and industry members should collaborate closely. The community should develop collaborations to advance an interoperable ecosystem of digital twins and share lessons learned with both the scientific community and the public. The strong interests and committed investments from national and intergovernmental organizations provide a unique opportunity for the community to work together to address the societal challenges that the world faces collectively.

## 5. Contributor Roles Taxonomy (CReDiT) Statement

YR: Conceptualization, Writing-Original draft, Writing-Reviewing and Editing; RR: Conceptualization, Writing-Original draft, Writing-Reviewing and Editing; KD: Conceptualization, Writing-Original draft, Writing-Reviewing and Editing; SEH: Conceptualization, Writing-Reviewing and Editing; AH: Conceptualization, Writing-Reviewing and Editing; AB: Conceptualization, Writing-Original draft, Writing-Reviewing and Editing; SB: Conceptualization, Writing-Original draft, Writing-Reviewing and Editing; TG: Conceptualization; DH: Conceptualization; BDS: Conceptualization; VR: Conceptualization; DN: Conceptualization, Writing-Original draft, Writing-Reviewing and Editing; EAK: Conceptualization, Writing-Reviewing and Editing.

## 6. Acknowledgement

Y. Rao is supported supported by NOAA through the Cooperative Institute for Satellite Earth System Studies under Cooperative Agreement NA19NES4320002. S.E. Haupt is with the National Center for Atmospheric Research, which is a major facility sponsored by the National Science Foundation under Cooperative Agreement No. 1852977. We want to thank the participants of the 4th NOAA AI Workshop contributing to the discussion around the topic of digital twins for Earth systems. We would also like to thank Barbara Ambrose for creating the figure based on an early draft, Lukas Noguchi for proofreading the manuscript, and Scott Cross for providing feedback to the manuscript.